\documentclass[12pt]{article}
\usepackage{bm}
\usepackage{a4wide,graphicx}
\usepackage{cite}
\usepackage{amssymb}
\usepackage{amsmath}
\usepackage{color}

\newcommand{\be}{\begin{equation}}
\newcommand{\ee}{\end{equation}}
\newcommand{\ba}{\begin{eqnarray}}
\newcommand{\ea}{\end{eqnarray}}

\begin{document}


\title{$\phi$ meson self-energy in nuclear matter from $\phi N$ resonant interactions}

\author{D.~Cabrera$^1$, A.~N.~Hiller Blin$^1$, M.~J.~Vicente Vacas$^1$\\
$^1$ Instituto de F\'{\i}sica Corpuscular, Universidad de Valencia--CSIC,\\
Institutos de Investigaci\'on, Ap. Correos 22085, E-46071 Valencia, Spain
}

\date{\today}

\maketitle
\begin{abstract}
The $\phi$-meson properties in cold nuclear matter are investigated by implementing resonant $\phi N$ interactions as described in effective approaches including the unitarization of scattering amplitudes. Several $N^*$-like states are dynamically generated in these models around $2$~GeV, in the vicinity of the $\phi N$ threshold. We find that both these states and the non-resonant part of the amplitude contribute sizably to the $\phi$ collisional self-energy at finite nuclear density. These contributions are of a similar strength as the widely studied medium effects from the $\bar K K$ cloud. Depending on model details (position of the resonances and strength of the coupling to $\phi N$) we report a $\phi$ broadening up to about $40$-$50$~MeV, to be added to the $\phi\to\bar K K$ in-medium decay width, and an attractive optical potential at threshold up to about $35$~MeV at normal matter density. The $\phi$ spectral function develops a double peak structure as a consequence of the mixing of resonance-hole modes with the $\phi$ quasi-particle peak. The former results point in the direction of making up for missing absorption as reported in $\phi$ nuclear production experiments.

\end{abstract}
\vskip 0.5 cm



\section{Introduction}  
\label{sec:intro}
The features of strongly interacting matter in a broad range of temperatures and densities has been a subject of great interest in the last decades, in connection with fundamental aspects of the strong interaction such as the nature of deconfinement or the physical mechanism of chiral symmetry restoration \cite{Rapp:1999ej}. Hadrons with strangeness, in particular, have been a matter of intense investigation regarding the study of exotic atoms \cite{Friedman:2007zza}, the analysis of strangeness production in heavy-ion collisions (HICs) \cite{Fuchs:2005zg,Forster:2007qk,Hartnack:2011cn} and the microscopic dynamics ruling the composition of neutron stars \cite{Kaplan:1986yq,Lattimer:2006xb}.

Vector-meson resonances with strange content, either open ($\bar K^*$, $K^∗$) or hidden ($\phi$), have triggered substantial attention recently, both theoretically and experimentally. An early motivation for present theoretical developments originates in the substantial experimental activity within the RHIC low energy scan program and the HADES experiment at GSI, which are currently performing measurements in order to extract the in-medium properties of hadronic resonances from HICs \cite{Blume:2011sb,Agakishiev:2013nta}.
A prominent example of the former is the production of $K^*$ mesons: Unlike their unflavored partners, strange vector mesons do not decay in the dilepton channel, making their experimental detection less clean. Still, there is hope that their in-medium properties can be accessed via reconstruction of their hadronic decays. Interestingly, the lack of baryonic resonances with strangeness $S = +1$ calls for much more moderate medium effects on the $K^*$ as compared to the $\bar K^*$, and thus a scenario of non-degenerate spectral functions for these vector mesons emerges in the presence of a baryonic medium.
Presently, there is also a big interest in the experimental study of production at ultrarelativistic energies since vector resonances with strange content are considered as a promising probe of the hot medium and the quark-gluon plasma (QGP) formation \cite{Rafelski:1982pu,SchaffnerBielich:1999uj,Kstar_probe,Ilner} (see also the recent study in Ref.~\cite{Cassing:2015owa} for an alternative interpretation of strangeness enhancement in HICs).

Interest in the $\phi$ meson properties has also been revived in recent years. The $\phi$, having a hidden strangeness content, strongly couples to the $\bar K K$ system and therefore its in-medium dynamics is highly governed by its decay into the light pseudoscalars, whereas its coupling to the nucleon is OZI-forbidden. The production of $\phi$ mesons in HICs is being investigated both at RHIC and at the LHC, with special attention to the ratio between dilepton and hadronic decays \cite{Wada:2013mua,Abelev:2014uua}.
Previously, the $\phi$ production in nuclei was studied by the KEK-PS-E325 Collaboration \cite{Muto:2005za}, the LEPS Collaboration at Spring8/Osaka \cite{Ishikawa:2004id}, the CLAS Collaboration at the Jefferson Lab \cite{Wood:2010ei}, and recently by the ANKE experiment at COSY-J\"ulich \cite{Hartmann:2012ia}. Together with theoretical analyses \cite{Cabrera:2003wb,Magas:2004eb,Hartmann:2012ia,Paryev:2008ck} on the $\phi$ nuclear transparency ratio it has been concluded that the $\phi$ experiences a huge absorption in nuclei, consistent with inelastic widths about one order of magnitude larger than its dominant decay in vacuum (e.g. in Ref.~\cite{Wood:2010ei} widths as large as $23-100$~MeV are reported for an effective nuclear density of $1/2\rho_0$, with $\rho_0$ the normal nuclear matter density\footnote{We recall that the extraction of the $\phi$ decay width -- or self-energy -- in the medium from observables such as the transparency ratio or decay spectra requires analyses via Glauber or transport model approaches. The CLAS results on the $\phi$ width, for example, rely on a Glauber model parameterization of the data with inelastic $\phi N$ cross sections in the range $16-70$~mb for different nuclei.}). These results point at medium effects typically larger than predicted by hadronic models, leaving room for the study of additional interaction processes with the nuclear medium.

Additional information on the $\phi$ dynamics in the medium has been revealed by recent measurements in HICs.
The observation of deep sub-threshold $\phi$ production by the HADES Collaboration in $1.23$~AGeV Au$+$Au collisions \cite{Lorenz:2014eja} calls for a review of possible interaction and/or production mechanisms that may have been missed in theoretical analyses.
A first attempt has been carried out in the context of the UrQMD transport model \cite{Steinheimer:2015sha} by implementing the production of $\phi$ mesons from the decay of massive $N^*$ resonances excited in initial $NN$ collisions (see also \cite{Moreau:2015ika,Cabrera:2015bga} for a study of the sensitivity of strange hadron observables to different production mechanisms, such as two-body hadronic processes, in the context of the Parton-Hadron-String Dynamics transport model).
Another mechanism leading to $\phi$ sub-threshold production relies on changes in the $\phi$ properties such as an attractive mass shift and/or an enhanced decay width as a consequence of $\phi N$ interactions. The large absorption advocated by the nuclear production experiments supports this possibility and encourages detailed studies on the $\phi$ properties in the nuclear medium, both at zero and finite temperature.

The $\phi$ self-energy in nuclear matter originating from medium effects on the $\bar K K$ decay mode has been profusely studied in the past, both at zero and finite temperature, taking advantage of chiral dynamics constraints on the kaon and antikaon interaction with nucleons and light mesons \cite{Ko:1992tp,Klingl:1997tm,Holt:2004tp,AlvarezRuso:2002ib,Oset:2000eg,Cabrera:2002hc,Faessler:2002qb}. The progress on unitarized approaches in coupled channels, based on the $SU(3)$ meson baryon chiral Lagrangian, confirmed the role of the $\Lambda (1405)$ as a leading actor responsible for some of the peculiar properties of the $\bar K N$ interaction at finite density. At the same time, in these approaches, the   $\Lambda (1405)$ appears as a dynamically generated resonance.
In our previous study~\cite{Cabrera:2002hc}, a $\phi$ width of about $28$~MeV and a very small attractive mass shift were obtained at normal nuclear matter density from $\phi$ decay processes mediated by the kaon cloud, such as $\phi N \to K Y$ and $\phi N \to [K \Lambda(1405)]\to K \pi Y$, including $s$- and $p$-wave $\bar K N$ interactions. Still, the large modification obtained for the $\phi$ width proved insufficient to describe the $\phi$ transparency ratio measured in proton induced and photoproduction experiments \cite{Ishikawa:2004id,Wood:2010ei,Hartmann:2012ia}. The recent measurements at COSY-ANKE have paved the missing information on the momentum dependence of the $\phi$ absorption below $1.5$~GeV$/c$ momenta (LEPS and CLAS measured transparency ratios for integrated average momenta of $1.8$ and $2.0$~GeV$/c$, respectively). Thus, theoretical calculations of the $\phi$ self-energy with special attention to its momentum dependence are highly motivated.

In the present work, in view of the recent theoretical and experimental developments, we revisit the $\phi$ self-energy in cold nuclear matter by analyzing the effect of direct $\phi$-nucleon interactions not previously considered. The idea relies on the fact that the hidden strangeness in the $\phi$ meson can be exchanged with the nucleon by the coupling to $K^*\Lambda$ and $K^* \Sigma$ pairs without violation of the OZI rule. This implies that elastic $\phi$ scattering proceeds via $K^*Y$ loops, which calls for an adequate treatment of unitarity in coupled channels (we recall that the evaluation of the self-energy requires the knowledge of the forward $\phi N$ elastic scattering amplitude).
In recent years two different coupled-channel approaches have been developed to assess the interaction of vector mesons with baryons. In Refs.~\cite{Oset:2009vf,Oset:2012ap} vector mesons are introduced within the Hidden Local Symmetry (HLS) approach and a tree-level vector-meson baryon $s$-wave scattering amplitude is derived as the low-energy limit of a vector-meson exchange mechanism. In Ref.~\cite{Gamermann:2011mq}, the interaction of vector mesons with  baryons is obtained on the basis of a $SU(6)$ spin-flavor symmetry extension of standard $SU(3)$ meson-baryon chiral perturbation theory, leading to a generalized ($s$-wave) Weinberg-Tomozawa interaction between pseudoscalar and vector mesons and the baryon octet/decuplet, while preserving the chiral symmetry results for the interactions of the pseudoscalar mesons.
In the $S=0$ sector, these two models share a crucial feature for our purpose. Several broad $N^*$ states are dynamically generated around a center-of-mass energy $\sqrt{s}\simeq 2$~GeV, immediately above the $\phi N$ threshold and with a non-negligible coupling to this channel, which may lead to a sizable $\phi$ self-energy contribution. It is the purpose of this work to explore this possibility and to study whether resonant $\phi N$ interactions can fill in for the missing absorption pointed out by nuclear production experiments.

Future experimental activity on nuclear production of vector mesons is to be carried out at the Japan Proton Accelerator Research Complex (J-PARC). Our present study is thus particularly timely.
With the focus on further applications in transport model simulations for HIC experiments such as the present HADES program and the forthcoming activity at GSI-FAIR and NICA-JINR, we present our results in terms of the $\phi$-meson nuclear optical potential and the (off-shell) spectral function. Momentum dependence is fully taken into account. Extension of our present calculation to a finite-temperature baryonic medium is straightforward and will be addressed in future work.
This letter is organized as follows: 
In Sec.~\ref{sec:model} we detail the calculation of the $\phi$ self-energy building on our previous results for the $\bar K K$ contribution by incorporating the additional $\phi N$ interaction mechanism within the HLS and $SU(6)$ approaches. Next, we present our results for the $\phi$ nuclear optical potential and spectral function in Sec.~\ref{sec:results}. Finally, we summarize and discuss possible applications of our results in Sec.~\ref{sec:summary}.

\section{$\phi$ self-energy}
\label{sec:model}
In the present study, we focus on $\phi$ medium effects in the light of recent developments on the meson baryon interaction in $SU(3)$ chiral effective theories and their extensions to include the octet (nonet) of light vector mesons within a consistent theoretical framework.
On top of the former, the $\phi$ is subject to a sizable renormalization in nuclear matter from its coupling to $\bar K K$, which is the dominant decay channel in vacuum (accounting for a branching ratio of about 83\% of the total decay width). To start with, we review here briefly the $\phi$ self-energy arising from its $\bar K K$ cloud both in vacuum and in nuclear matter, where the antikaons experience dramatic effects resulting from their resonant scattering with the nucleons.

The coupling of the $\phi$ to pseudoscalar mesons has been thoroughly studied and we take recourse of the classical references \cite{Schechter:1986vs,Jain:1987sz,Meissner:1987ge,Klingl:1996by} in the context of non-linear realizations of chiral symmetry for the pseudoscalar sector (i.e. chiral perturbation theory), where the vector fields are gauged in the theory in analogy to the electromagnetic current, automatically fulfilling the requirements of vector-meson dominance. An alternative way to introduce vector mesons in compliance with chiral symmetry constraints was developed in the HLS approach  \cite{Bando:1984ej}, which has been recently employed with success to generate the interactions of vector mesons with baryons from the octet ($p$, $n$) and the decuplet ($\Delta$).

The interaction Lagrangian providing the coupling of the $\phi$ field to $K$ and $\bar K$ reads
\begin{eqnarray} \label{eq:Lagr-phi-KKbar}
{\cal L}_{\phi \bar K K} &=& -i g_{\phi} \phi_{\mu} (K^-
\partial^{\mu} K^+ - K^+ \partial^{\mu} K^- + \bar{K}^0
\partial^{\mu} K^0 - K^0 \partial^{\mu} \bar{K}^0)
\nonumber \\
&+& g_{\phi}^2 \phi_{\mu} \phi^{\mu} (K^- K^+ + \bar{K}^0 K^0) ,
\end{eqnarray}
with notation and $SU(3)$ field definitions as in Ref.~\cite{Oset:2000eg}. At leading order the $\phi$ self-energy gets contributions from $\bar K K$ loop diagrams and tadpole diagrams. The computation is straightforward and one finds
\begin{equation}
\label{eq:phi_self_vac}
-i \Pi_{\phi}^{\bar K K} (q) = 2 
g_{\phi}^2 \frac{4}{3} \int  \frac{d^4k}{(2\pi)^4} 
\bigg \lbrack \frac{(q \cdot k)^2}{q^2}-k^2 \bigg \rbrack
D_K(q-k)
D_{\bar{K}}(k)
+
2 g_{\phi}^2 \int  \frac{d^4k}{(2\pi)^4} \lbrack D_K(k)+ D_{\bar{K}}(k) \rbrack
\end{equation}
for a $\phi$ meson in vacuum with momentum $q=(q^0,\vec{q})$, and $D_{K(\bar K)}$ representing the kaon (antikaon) propagator. The $\phi$ decay width to $\bar K K$ can be obtained from the imaginary part of the self-energy, namely
\begin{equation}
\label{eq:phi_self_Im}
\textrm{Im}\, \Pi_{\phi}^{\bar K K}(q^2) = -\frac{g_{\phi}^2}{24\pi \sqrt{q^2}} 
(q^2-4 m_K^2)^{\frac{3}{2}} \ \Theta(q^2-4 m_K^2) \ .
\end{equation}
At rest one has $\Gamma_{\phi}=-\textrm{Im}\, \Pi_{\phi}^{\bar K K}(q^0=M_{\phi},\vec{q}=\vec{0})/M_{\phi}=\frac{g_{\phi}^2}{24\pi} |\vec{k}_{cm}|^3/M_{\phi}^2$, which sets the value of the $\phi~\bar K K$ coupling constant to $g_{\phi}=4.57$ so as to reproduce $\Gamma^{\textrm{exp}}_{\phi\to \bar K K} = 3.54$~MeV, corresponding to a branching ratio $BR(\phi\to \bar K K)=83\%$ of the total decay width \cite{Agashe:2014kda}. Here, $M_{\phi}$ and $m_K$ are the $\phi$ and $K$ mass, respectively.

\subsection{Contribution from the in-medium $\bar K K$ cloud}
\label{subsec:KKbar}
Light vector mesons decay strongly into pseudoscalars. This is a prominent source of medium effects as these pseudoscalars (e.g. $\rho\to\pi\pi$, $\omega\to 3\pi$) also modify their properties sizably in the nuclear medium. In the present case, it is the nuclear dynamics of the kaons and antikaons which plays an important role, particularly due to the sub-threshold excitation of the $\Lambda(1405)$ resonance in $s$-wave $\bar K N$ scattering as well as other deeply sub-threshold hyperons [$\Lambda$, $\Sigma$ and $\Sigma^*(1385)$] in the $p$ wave.
These effects are incorporated in the many-body theoretical framework by dressing the kaon (antikaon) propagator with the self-energy originating from the $KN$ ($\bar K N$) interaction. The complicated analytical structure of in-medium propagators calls for the use of a spectral (Lehmann) representation, which simplifies notably the calculation of the vector-meson self-energy. The in-medium kaon propagator thus reads
\be
\label{eq:Lehmann-rep}
D_{\bar{K} (K)} (q^0,\vec{q};\rho) =
\int_0^{\infty} d\omega \bigg ( \frac{S_{\bar{K} (K)}
(\omega,\vec{q};\rho)}{q^0-\omega+i\eta} - \frac{S_{K (\bar{K})}
(\omega,\vec{q};\rho)}{q^0+\omega-i\eta} \bigg ) \ ,
\ee
where $S_{\bar{K} (K)}$ is the spectral function of the $\bar{K} (K)$ meson,
\be
\label{eq:spectral-kaons}
S_{\bar{K} (K)}(q^0,\vec{q};\rho) = - \frac{1}{\pi}
\frac{\textrm{Im} \, \Pi_{\bar{K} (K)}(q^0,\vec{q};\rho)}
{| (q^0)^2 - \vec{q}\,^2 - m_K^2 - \Pi_{\bar{K} (K)}(q^0,\vec{q};\rho)|^2} \ ,
\ee
with $\Pi_{\bar{K} (K)}$ the antikaon (kaon) self-energy.
Using the previous expression in Eq.~(\ref{eq:phi_self_vac}) allows to evaluate the energy integration analytically. Up to a pole-free integration over kaon momenta $k$ and angle $\theta$ with respect to the $\phi$ momentum, the imaginary part of the $\phi$ self-energy from the $\bar K K$ cloud is given by
\begin{eqnarray}
\label{eq:Im-phi-self-KKbar}
\textrm{Im} \, \Pi_{\phi}^{\bar{K}K} (q^0,\vec{q};\rho)
&=& - \frac{1}{4\pi} g_{\phi}^2 \frac{4}{3} \int_0^{\infty} dk \, \vec{k}\,^2
\int_{-1}^{1} d(\cos{\theta}) \frac{1}{\widetilde{\omega}(q-k)} \times
\nonumber \\
& &
\!\!\!\!\!\!\!\!\!\!\!\!\!\!\!\!\!\!\!\!
\bigg \lbrack \frac{(q \cdot k)^2}{M_{\phi}^2}-k^2 \bigg
 \rbrack_{k^0=q^0-\widetilde{\omega}(q-k)}
S_{\bar{K}}(q^0-\widetilde{\omega}(q-k),\vec{k};\rho) \,
\Theta (q^0-\widetilde{\omega}(q-k))
\ ,
\end{eqnarray}
where $\widetilde{\omega}(q-k)=[(\vec q- \vec k)^2+m_K^2+\Pi_K(\rho)]^{1/2}$. In the former result, we have approximated the $K$ spectral function by a delta function with a quasi-free dispersion relation accounting for the kaon self-energy, which we calculate within a $T\rho$ approximation as is justified below. The imaginary part of the $\phi$ self-energy encodes the enhancement of the $\phi$ decay width in a nuclear environment or, in other words, is related to the $\phi$-nucleus inelastic (absorptive) cross section. It therefore plays an essential role to estimate the $\phi$ transparency ratio in nuclear production experiments \cite{Cabrera:2003wb,Magas:2004eb,Hartmann:2012ia}.

The real part of the self-energy modifies the $\phi$ dispersion relation and can be interpreted as a density-dependent mass shift for $\phi$ mesons at rest in the nuclear matter reference frame. We note, however, that the self-energy has an explicit dependence on the $\phi$ energy and momentum. In realistic scenarios the $\phi$ is produced in the medium with a certain momentum distribution, thus it is convenient to keep the momentum dependence of the $\phi$ self-energy and spectral function. For certain applications, such as the propagation of hadrons in transport simulations of heavy-ion collisions or the production of mesic atoms and nuclei, the momentum-dependent nuclear optical potential is of use. This is defined for the $\phi$ in terms of the self-energy at the energy given by the in-medium dispersion equation, namely $V_{\phi}(q,\rho) = \Pi_{\phi}(\tilde{\omega}_{\phi}(q),\vec{q};\rho)/(2~\tilde{\omega}_{\phi}(q))$, where $\tilde{\omega}_{\phi}$ satisfies $(q^0)^2-\vec{q}\,^2-M_{\phi}^2-{\rm Re}\,\Pi_{\phi}(q^0,\vec{q};\rho)=0$ with $q^0=\tilde{\omega}_{\phi}$ (energy of the $\phi$ quasi-particle mode).
The integrals in Eq.~(\ref{eq:phi_self_vac}) are divergent for the real part. As done in Ref.~\cite{Cabrera:2002hc,Tolos:2010fq} we regularize our result by subtracting the vacuum $\phi$ self-energy. Alternatively, a dispersion relation over the imaginary part of the self-energy can be performed \cite{Klingl:1997kf}.

The $\bar K$ and $K$ self-energies in symmetric nuclear matter have been addressed within a chiral unitary approach to the kaon-nucleon interaction. Details can be found, for instance, in Ref.~\cite{Tolos:2006ny} and references therein (for nuclear matter studies at finite temperature see e.g.~\cite{Tolos:2008di,Cabrera:2014lca}).
In our model, we account for the kaon-nucleon interaction in coupled channels, including both $s$- and $p$-waves. The following channels, completing a $SU(3)$ basis of meson-baryon states, are taken into account: in the $S=-1$ sector one has, in the physical (charge) basis, $K^- p$, $\bar{K}^0n$, $\pi^0
\Lambda$, $\pi^0 \Sigma^0$, $\eta \Lambda$, $\eta \Sigma^0$, $\pi^+
\Sigma^-$, $\pi^- \Sigma^+$, $K^+ \Xi^-$, and $K^0 \Xi^0$. Since we restrict our study for symmetric nuclear matter it is convenient to work in the isospin basis ($I=0,1$). For obvious reasons the $S=+1$ sector is elastic with one single channel, $KN$.
The tree-level $s$-wave amplitudes are fixed by chiral symmetry breaking and given by the lowest order (LO) Weinberg-Tomozawa chiral Lagrangian describing the interaction between the pseudoscalar meson octet and the $J^P=1/2^+$ baryon octet. Unitarization along the right-hand cut in coupled channels is achieved by solving the Bethe-Salpeter equation with on-shell amplitudes, schematically $T=V+VGT$, extending the applicability of the low-energy theory.
Here $V$ contains the LO meson-baryon amplitude and $G$ stands for the meson-baryon resolvent or loop function.
The formalism is modified in the medium to
account for Pauli blocking effects, mean-field binding potentials on
baryons, and pion and kaon self-energies, which are implemented in the meson-baryon loop functions\footnote{The present model has been improved recently to account for
unitarization and self-consistency for both the $s$- and $p$-wave interactions
at finite temperature and density \cite{Cabrera:2014lca}, thus extending its applicability to the physics of heavy-ion collisions.}.
$G$ depends on the $\bar K$ self-energy through the meson propagator, which at the same time is evaluated by summing the effective $\bar K N$ interaction over the nucleon Fermi distribution, schematically $\Pi_{\bar K(K)}=\sum_{\vec{p}}n(\vec{p}\,) T_{\bar K(K)N}$. The former thus leads to a self-consistent solution of both the kaon-nucleon $T$-matrix and the kaon self-energy in nuclear matter.

The isoscalar, $s$-wave $\bar K N$ amplitude is dominated by the excitation of the $\Lambda(1405)$ right below threshold, which acquires  its physical width dominantly from the decay into $\pi \Sigma$ states. When the nuclear medium is switched on, the resonance  is practically washed out and its strength spread out over energy, as a consequence of the in-medium decay mechanisms incorporated through the self-consistent dressing of mesons, e.g. $\Lambda(1405) \to \pi (YN^{-1}) N, \pi (NN^{-1})\Sigma, \pi (\Delta N^{-1})\Sigma$. The former modifications lead to an attractive $\bar K$ optical potential of $-60$~MeV at normal nuclear matter density for antikaons at rest, and a largely spread $\bar K$ spectral function. At finite momentum the $p$-wave self-energy, built up from the coupling to $\Lambda(1115)N^{-1}$, $\Sigma(1195)N^{-1}$, and $\Sigma^*(1385)N^{-1}$ excitations, contributes substantially by populating the low-energy region of the $\bar K$ spectrum.
Due to the absence of $S=+1$ baryonic resonances, the $KN$ amplitude is mildly energy dependent and to a good approximation the $K$ self-energy can be calculated in the $T\rho$ form. The kaons experience a repulsive mass shift of about $30$~MeV at $\rho=\rho_0$ whereas their spectral function stays narrow in the medium and only at high temperatures some thermal broadening can be appreciated \cite{Tolos:2008di}. In Eq.~(\ref{eq:Im-phi-self-KKbar}) the $K$ self-energy is given by $\Pi_K(\rho) = 0.13 \,m_K^2 \rho / \rho_0$.

Finally, it is worth mentioning that once meson-baryon interactions are considered in the theory, vertex corrections are demanded by gauge invariance. These were considered in full in Refs.~\cite{Klingl:1997tm,Cabrera:2002hc} for the $\phi$ meson at rest and extended to the case of finite momentum in Ref.~\cite{Cabrera:2003wb}. We refer the reader to the former references for detailed explanations.

\subsection{Contribution from resonant meson-baryon interactions}
\label{subsec:phi-N}
In this section we discuss a novel contribution to the $\phi$ self-energy emerging from vector-meson baryon interactions as it has been addressed recently in chirally motivated theories implementing unitarization in coupled channels.
The properties of vector mesons within these approaches have already been analyzed for the case of the $K^*(892)$, which, despite the absence of a dileptonic decay mode, making its experimental detection less clean, has triggered much attention lately and is currently been investigated in heavy-ion collision reactions at low (HADES at GSI/FAIR) and ultrarelativistic energies (RHIC Beam Energy Scan). The self-energy of the $\omega$ meson was also updated along the same ideas in Ref.~\cite{Ramos:2013mda}.

Our study of the $\phi$ self-energy contribution from vector-meson baryon interactions follows the same strategy and we build upon the results of Refs.~\cite{Oset:2009vf} and \cite{Gamermann:2011mq}.
In Ref.~\cite{Oset:2009vf} the interaction between the octet of light vector mesons (plus the $SU(3)$ singlet) and the octet of $J^P=1/2^+$ baryons was studied within the HLS formalism, which allows to incorporate vector-meson interactions with pseudoscalars respecting the chiral dynamics of the pseudoscalar meson sector. The interactions with baryons are assumed to be dominated by vector-meson exchange diagrams, which allows for an interpretation of chiral Lagrangians in the meson-baryon sector as the low-energy limit of vector-exchange mechanisms naturally occurring in the theory. Vector-meson baryon amplitudes emerge in this scheme at leading order, in complete analogy to the pseudoscalar-meson baryon case, via the self-interactions of vector-meson fields in the HLS approach.
At small momentum transfer, the LO $s$-wave scattering amplitudes have the same analytical structure as the ones in the pseudoscalar meson baryon sector (Weinberg-Tomozawa interaction), namely
\begin{eqnarray}
\label{eq:VB-potential}
V_{ij}
&=& - C_{ij} \frac{1}{4f^2} (2\sqrt{s}-M_{B_i}-M_{B_j}) \,
\left ( \frac{M_{B_i}+E_i}{2 M_{B_i}} \right )^{1/2} \left ( \frac{M_{B_j}+E_j}{2 M_{B_j}} \right )^{1/2}  \,\vec{\varepsilon}\cdot\vec{\varepsilon}\,' \nonumber \\
&\simeq& - C_{ij} \frac{1}{4f^2}(q^0+q'^0) \,\vec{\varepsilon}\cdot\vec{\varepsilon}\,' \ ,
\end{eqnarray}
where $E_i$ and $M_{B_i}$ denote the center-of-mass energy and mass of the baryon in channel $i$, $f$ stands for the meson decay constant, $C_{ij}$ are the channel-dependent symmetry coefficients \cite{Oset:2009vf}, and the Latin indices label a specific vector-meson baryon channel, e.g. $\phi p$. For practical purposes the second equation is satisfied to a good approximation, where $q^0$~($q'^0$) is the energy of the incoming~(outgoing) vector meson with polarization $\vec{\varepsilon}$~($\vec{\varepsilon}\,'$). The coupled-channel problem is conveniently organized in the isospin-strangeness basis with $(I,S)=(1/2,0)$, $(0,-1)$, $(1,-1)$, $(1/2,-2)$, corresponding to sectors where the LO interaction is attractive. Our interest is focused in the $I=1/2$ and $S=0$ sector, where the model comprehends the set of channels $\rho N$, $\omega N$, $\phi N$, $K^* \Lambda$, and $K^* \Sigma$.
Upon unitarization, several meson-baryon resonances are dynamically generated which in some cases can be assigned to existing experimental observations whereas in other cases constitute predictions for new states. In the $\phi N$ channel, it turns out that a broad resonance is generated around $\sqrt{s}=2$~GeV, immediately above the $\phi N$ threshold, thus potentially contributing to the $\phi$ collisional self-energy. Also, it will clearly mix with the $\phi$ quasi-particle mode in the $\phi$ spectral function. By means of a residue analysis of the $T$-matrix pole corresponding to this state it was concluded that it mostly couples to $K^*Y$ and much less strongly to the $\phi N$ and $\omega N$ channels. However, as we discuss in the next section, the interaction in the $\phi N$ channel is sufficiently strong to generate a substantial contribution to the $\phi$ self-energy, competing in size with (and coherently summing to) the self-energy from the $\bar K K$ cloud.

In Ref.~\cite{Gamermann:2011mq}, an extension of the standard $SU(3)$ meson-baryon chiral Lagrangian based on $SU(6)$ spin-flavor symmetry was devised in order to analyze the low-energy interactions between pseudoscalar and vector mesons with the baryon octet ($N$) and decuplet ($\Delta$).
The model comprehends several sectors of total spin, isospin and strangeness (with negative parity)\footnote{The LO $s$-wave amplitudes in the $SU(6)$ model have the same form as in Eq.~(\ref{eq:VB-potential}), with different symmetry coefficients, channel-dependent meson decay constants, and a larger coupled-channel space.}.
Once unitarization in coupled channels is implemented on the $s$-wave amplitudes, a number of baryonic resonances are dynamically generated in the energy range from the $N^*(1535)$ up to $\sim$2.2~GeV. In particular one finds that two resonances slightly above the $\phi N$ threshold emerge in this approach, with $J^P=1/2^-,3/2^-$.
These two states appear in the SU(6) model with slightly different masses  to the one observed in the  HLS amplitudes of Ref.~\cite{Oset:2009vf} and are also found to couple strongly to $K^*\Lambda$ and $K^*\Sigma$.
The emergence of resonances in the region near $2$~GeV from vector-meson baryon interactions has been found crucial to correctly interpret the data shape and energy dependence of the $\gamma p\to K^0\Sigma^+$ reaction by the CBELSA/TAPS Collaboration \cite{Ramos:2013wua}. Claims for the need of a $N^*$ resonant state in that energy region have also been made from the analysis of $\Lambda(1520)$ photoproduction data by the LEPS Collaboration~\cite{Ryu:2016jmv}.

The two approaches discussed above predict relevant phenomenology for the modification of the $\phi$ properties in the medium, namely the presence of resonant strength in the $\phi N$ scattering amplitude in the vicinity of the $\phi N$ threshold. Nevertheless, the amplitudes in each model differ in size and energy dependence and we deem appropriate to perform our calculation of the $\phi$ self-energy in both approaches to assess the level of uncertainty from model dependence.
Within the $S=0$ sector let us denote by $T_{\phi N}^{IJ}$ the $\phi N$ scattering amplitude in a given isospin and total angular momentum channel (the $J$ index is irrelevant in the approach of Ref.~\cite{Oset:2009vf}), as a result of solving the coupled-channel Bethe-Salpeter equation in the on-shell factorization, $T = [1-VG]^{-1} V$, with the usual meaning for $V$ and $G$. The contribution to the $\phi$ self-energy proceeds by summing the scattering amplitude over the nucleon Fermi distribution,
\begin{equation}
\label{eq:phi-N-self}
\Pi^{\phi N}_{\phi}(q^0,\vec{q};\rho) =
4 \int \frac{d^3 p}{(2\pi)^3} n(\vec{p}) \frac{1}{6}[2 T^{\frac{1}{2}\frac{1}{2}}_{\phi N}(P^0,\vec{P})+4 T^{\frac{1}{2}\frac{3}{2}}_{\phi N}(P^0,\vec{P})] \ ,
\end{equation}
where $P^0=q^0+E_N(\vec{p})$ and $\vec{P}=\vec{q}+\vec{p}$ are the total energy and momentum of the $\phi N$ pair in the nuclear matter rest frame. The average over $J$ in the previous expression is trivial for the HLS amplitudes.

This study being the first approach to the $\phi$ self-energy from this sort of interaction models, we restrict ourselves to the following medium modifications: (i) $\phi$ self-energy from Eq.~(\ref{eq:phi-N-self}) with vacuum $\phi N$ scattering amplitudes (low-density approximation), and (ii) modification of the $\phi N$ $T$-matrix from Pauli blocking. Additional sources of medium effects include binding potentials for the baryons and the dressing of the $\phi$ propagator within the $\phi N$ loop function, leading to a self-consistent solution of both $\Pi^{\phi N}_{\phi}$ and $T_{\phi N}$. Self-consistency has been found to be important to account for when resonant states appear close to reaction thresholds, as is the case of the $\Lambda(1405)$ in the $\bar K N$ system. In the present case, one may expect a less severe effect since the relevant states couple mostly to $K^*Y$ channels. In any case an explicit evaluation is called for and, given the exploratory character of the present work, we relegate such a study for a follow-up publication.

Pauli blocking is straightforwardly implemented in the vector-meson baryon amplitudes by replacing the nucleon propagators in the loop function $G$ by the corresponding single-particle propagators in the Fermi gas approximation. At this level the in-medium $G$ function can be explicitly separated in a vacuum contribution plus a Pauli-blocking correction term,
\begin{eqnarray}
\label{eq:Pauli}
G(P;\rho) &=& G(P) + \delta G^{\rm Pauli}(P,\rho) \ , \nonumber \\
\delta G^{\rm Pauli}(P,\rho) &=& \int \frac{d^3 q}{(2\pi)^3} \frac{M_N}{E_N(\vec{p})} \frac{-n(\vec{p})}{[P^0-E_N(\vec{p})]^2-\omega_{\phi}^2(\vec{q})+i\epsilon}\bigg |_{\vec{p}=\vec{P}-\vec{q}} \ ,
\end{eqnarray}
with $\omega^2_{\phi}(\vec{q})=M_{\phi}^2+\vec{q}\,^2$ and $n(\vec{p})$ the Fermi gas nucleon momentum distribution, $n(\vec{p})=\Theta(p_F-|\vec{p}|)$, with $p_F=(3\pi^2\rho/2)^{1/3}$ the Fermi momentum in terms of the nuclear matter density $\rho$.
The integrals in $\delta G^{\rm Pauli}$ are finite whereas the vacuum $G$ is regularized dimensionally or in terms of an equivalent momentum cut-off \cite{Oset:2009vf,Gamermann:2011mq,Tolos:2010fq}.
We note that the Pauli-blocking correction only affects the meson-baryon loop function for channels containing nucleons.

\section{$\phi$ optical potential and spectral function in the nuclear medium}
\label{sec:results}

\begin{figure}[t]
\centering
\includegraphics[width=0.45\textwidth]{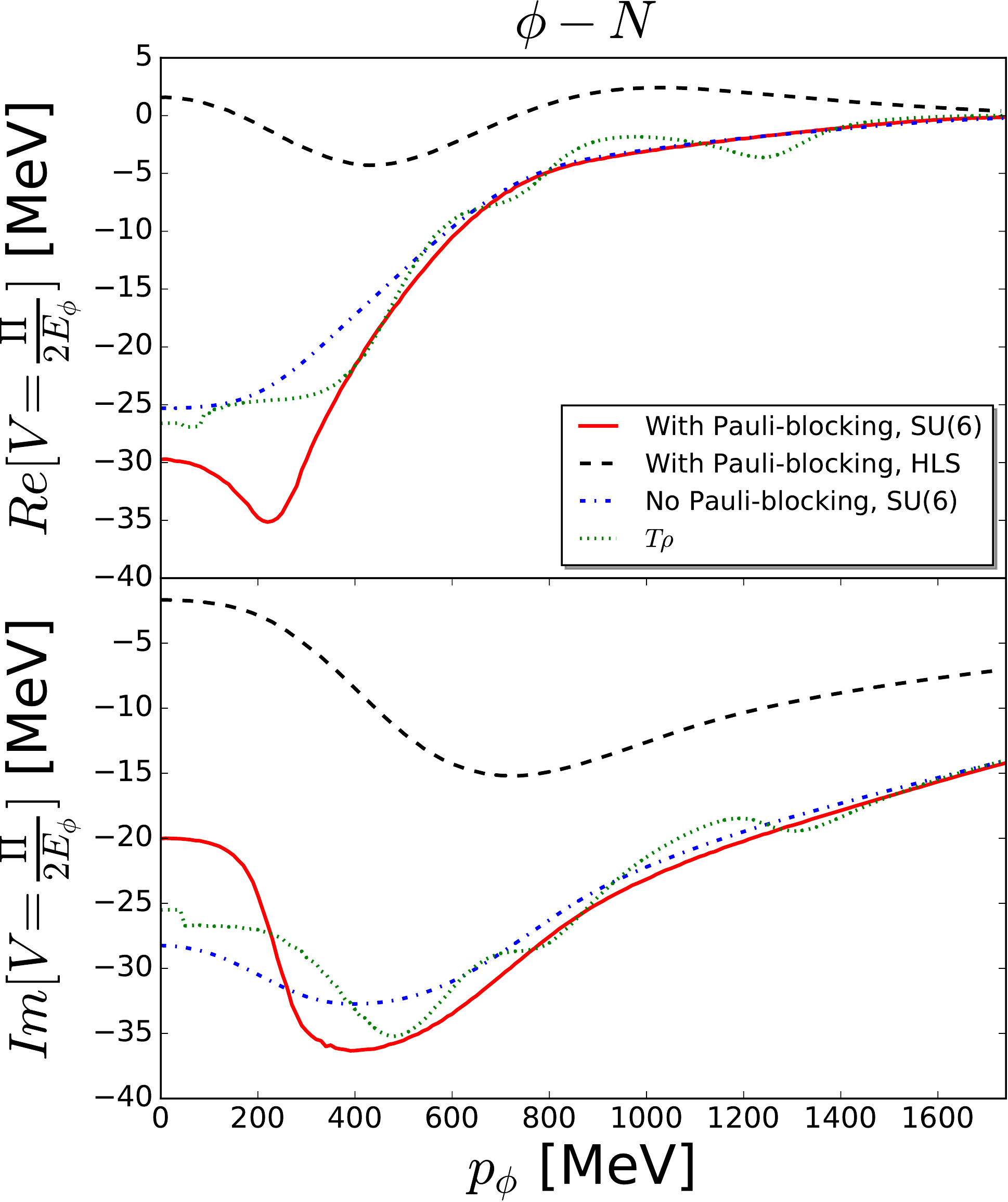} \hspace{10mm}
\includegraphics[width=0.45\textwidth]{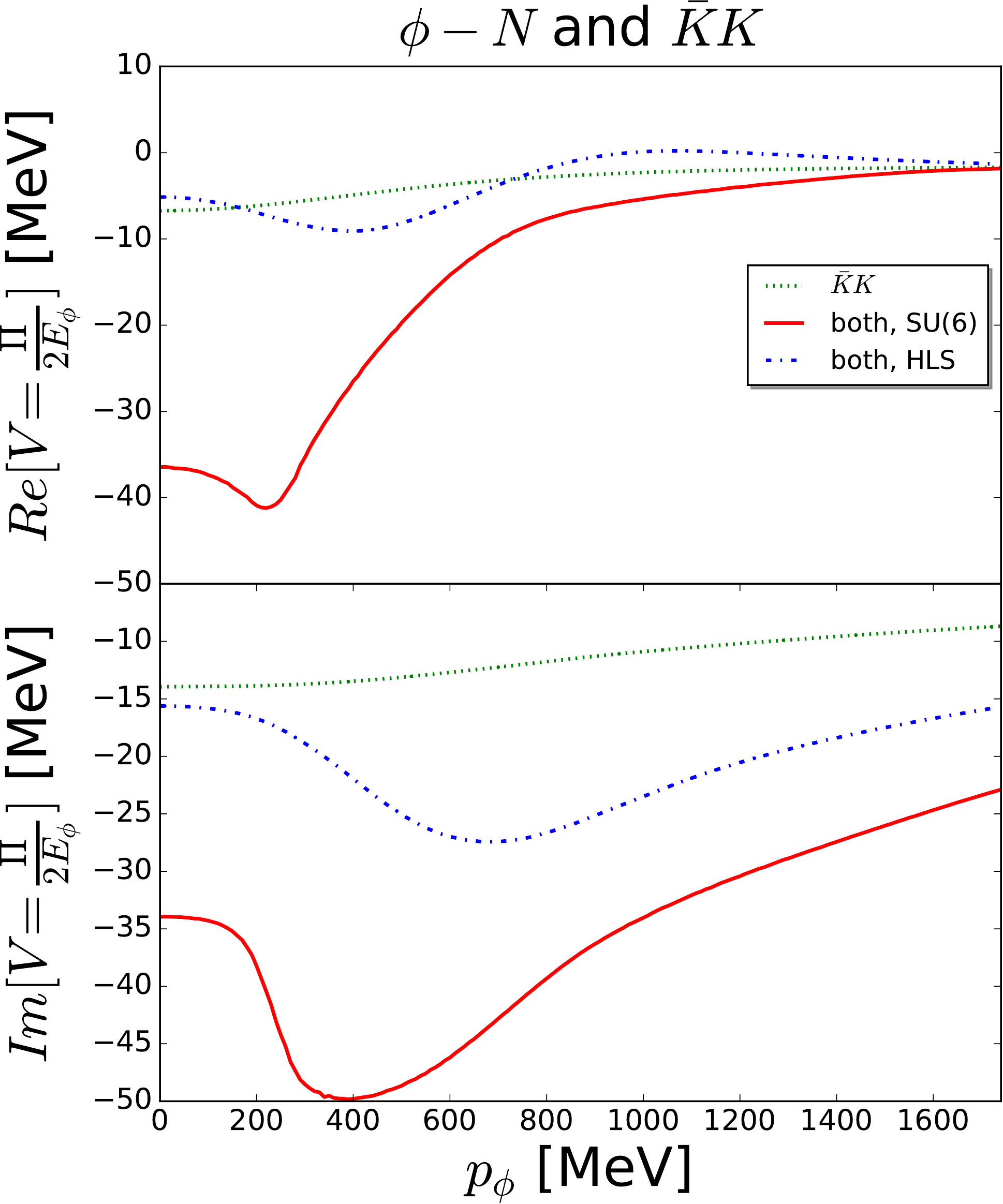}
\caption{Real and imaginary parts of the $\phi$ optical potential as a function of the $\phi$ momentum at normal matter density. Left: $\phi N$ scattering contribution from the $SU(6)$ chiral approach in a $T\rho$ approximation (dotted), using the self-energy evaluation in Eq.~(\ref{eq:phi-N-self}) with vacuum amplitudes (dash-dotted), and including on top Pauli-blocking effects (solid). The dashed curve corresponds to the evaluation with the $T$-matrix from the HLS approach. Right: Contribution from the $\bar K K$ cloud (dotted) and the coherent sum including $\bar K K$ plus $\phi N$ effects in the $SU(6)$ (solid) and HLS (dash-dotted) approaches.}
\label{fig:Vopt}
\end{figure}

\begin{figure}[t]
\centering
\includegraphics[width=0.45\textwidth]{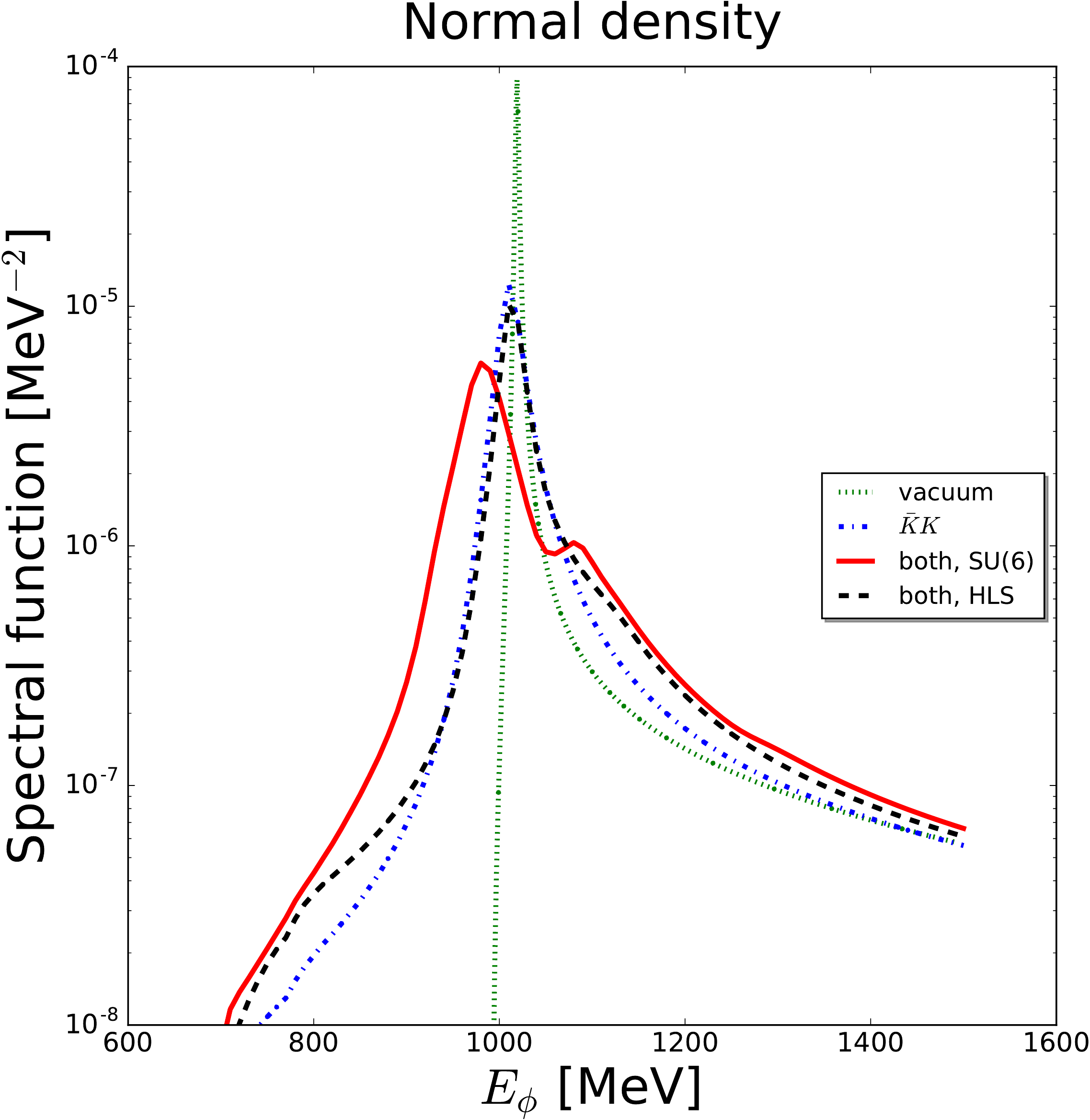} \hspace{10mm}
\includegraphics[width=0.45\textwidth]{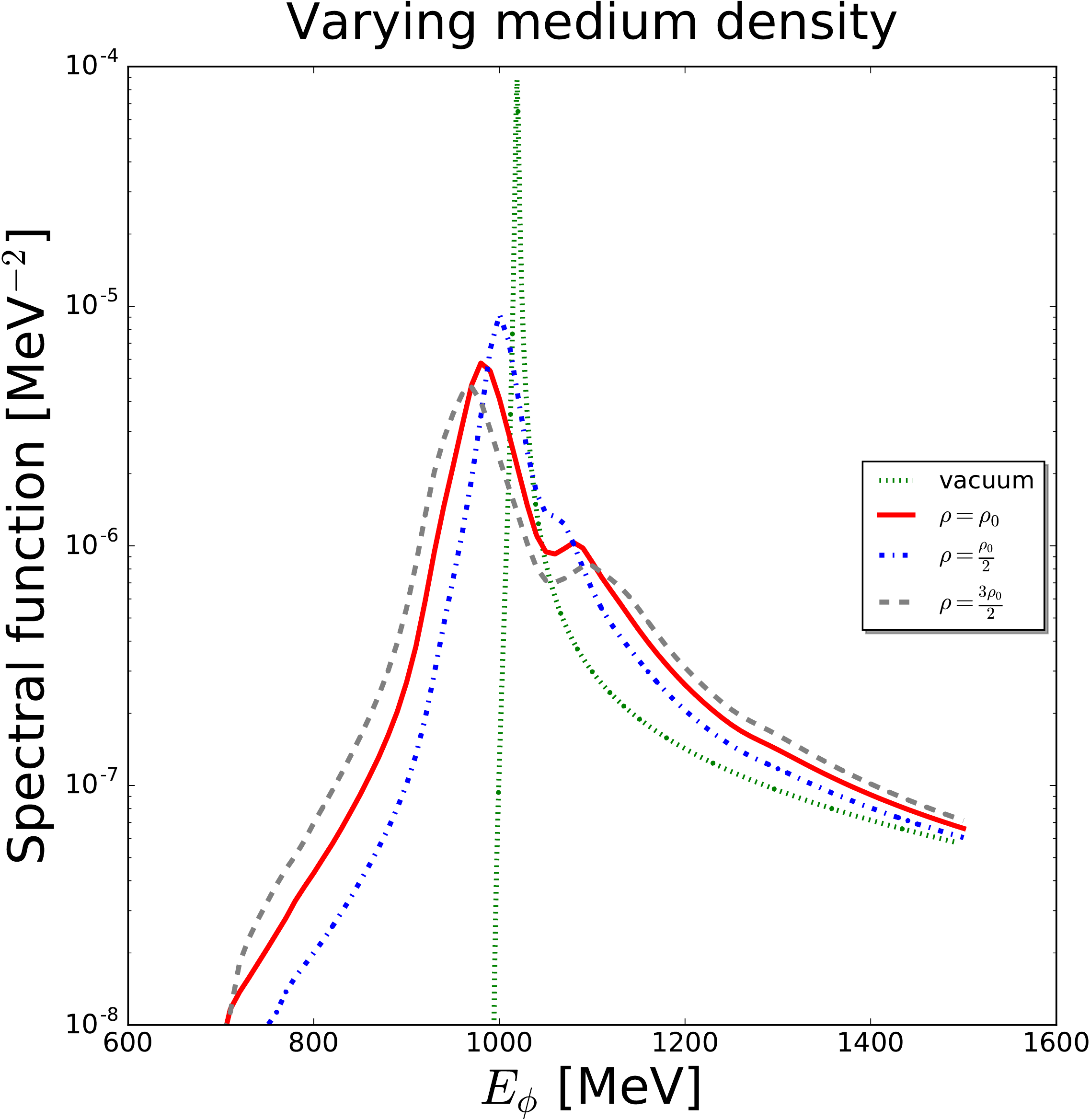} 
\caption{$\phi$ spectral function as a function of the (off-shell) $\phi$ energy at zero momentum. Left: calculation accounting for the $\bar K K$ self-energy (dash-dotted) or the $\bar K K$ plus $\phi N$ self-energies in the HLS model (dashed) and the $SU(6)$ model (solid), all three cases considered at $\rho=\rho_0$. The $\phi$ spectral function in vacuum is also shown (dotted). Right: Evolution with the nuclear density with the full model in the $SU(6)$ evaluation. Note that the solid and the dotted curves are the same in both plots.}
\label{fig:Spec-contrib}
\end{figure}

We start this section with the focus on medium effects arising from vector-meson baryon scattering. A quick estimation of the $\phi$ collisional width generated by this mechanism can be readily obtained from the total $\phi N$ cross section, $\Gamma_{\phi}^{\phi N}=\sigma_{\phi N} v_{\phi} \rho$, with $v_{\phi}=p_{\phi}/\omega_{\phi}$ in the Lab frame, resulting in values of tens of MeV for $\phi$ momenta below $1$~GeV$/c$. This result encourages more detailed calculations as discussed in the previous section.

In Fig.~\ref{fig:Vopt}, we show the real and imaginary parts of the $\phi$ nuclear optical potential as a function of the $\phi$ momentum, $p_\phi$. In the left panel, we depict the contribution to $V_{\phi}(p_\phi,\rho)$ coming from the $\phi N$ interaction at normal matter density for different evaluations of the $\phi$ self-energy using the scattering amplitudes from the $SU(6)$ model in Ref.~\cite{Gamermann:2011mq}:  $T\rho$ approximation (dotted), Eq.~(\ref{eq:phi-N-self}) with vacuum amplitudes and Fermi motion (dash-dotted), and the same including Pauli-blocking effects in the amplitude (solid). Moreover, we include the result with Pauli blocking for the calculation using the scattering amplitudes from \cite{Ramos:2013wua} (dashed), which are based on the HLS approach in Ref.~\cite{Oset:2009vf} with further constraints on the model from experimental data on the $\gamma p \to K^0 \Sigma^+$ reaction in the $K^*\Lambda$ region.
In the $SU(6)$ model, we observe that the $\phi N$ interaction induces an attractive potential of $-30$~MeV at rest with a considerable energy dependence when increasing the $\phi$ momentum, which can be traced back to the resonant states excited immediately above the $\phi N$ threshold. The imaginary part of the potential starts at $-20$~MeV and quickly raises beyond $q\sim 200$~MeV$/c$ reaching an extremum of $-35$~MeV and decreasing in magnitude thereon following a typical phase-space behavior.
We find that the $T\rho$ approximation is in reasonable agreement with Eq.~(\ref{eq:phi-N-self}) when vacuum amplitudes are considered, the latter exhibiting an energy smearing from the Fermi motion of nucleons, as expected.
However, when Pauli blocking is switched on, the optical potential deviates from the low-density approximation, particularly at $\phi$ momenta below $600$~MeV$/c$, where the typical nucleon momentum is of the order of the Fermi momentum. The attraction at vanishing $\phi$ momentum is increased by $5$~MeV whereas the collisional width is reduced by about $30$\% (recall that $\Gamma_{\phi}=-2\,{\rm Im}V_{\phi}$).

We note that the same calculation with the $T$-matrix from the HLS approach provides a much weaker self-energy at small $\phi$ momenta and a milder momentum dependence. The imaginary part of $V_{\phi}$ grows slowly in this case and reaches an extremum of about $-15$~MeV at $p_\phi\simeq 700$~MeV$/c$, clearly reflecting the position of the resonance in this model.
The differences in the optical potentials obtained for the two approaches considered in this work can be attributed to the following reasons: first, the $SU(6)$ model exhibits several resonant states in the region $\sqrt{s}\sim 2$~GeV, some of them with a high multiplicity since both $J=1/2,3/2$ channels contribute to the self-energy, whereas in the HLS amplitudes one single state at about $2.08$~GeV is found; and second, the behavior of the $\phi N$ amplitudes at threshold in both models is substantially different: whereas in the HLS the amplitude is small and mildly energy dependent below the resonant state (as expected from an $s$-wave interaction), the $SU(6)$ amplitude exhibits a pronounced peak at threshold which quickly falls off with energy in the $J=1/2$ channel. Such behavior corresponds to a would-be resonant state (``failed state'') whose strength starts to accumulate below threshold but is abruptly cut down by the opening of the $\phi N$ channel (Flatt\'e effect). The strength of this peak in the amplitude, not directly related to the excitation of a resonance (a pole of the scattering amplitude), determines the leading contribution to the self-energy at threshold.

In the right panel of Fig.~\ref{fig:Vopt} we show the contributions to the $\phi$ nuclear optical potential from the two sources of medium effects discussed in Sec.~\ref{sec:model}, namely the $\bar K K$ decay mode and the $\phi N$ interaction, at $\rho=\rho_0$. Both mechanisms give rise to optical potentials of the same order of magnitude and lead to a total potential of $(-36-34\,i)$~MeV at zero $\phi$ momentum when the $SU(6)$ $\phi N$ interaction is used, and of $(-5-16\,i)$~MeV when HLS amplitudes are considered. We should remark that even the much weaker effects obtained for the HLS model substantially modify the $\phi$ width for momenta around 500~MeV/c, and thus are relevant for the interpretation of typical production experiments.

Finally, we depict in Fig.~\ref{fig:Spec-contrib} the (off-shell) $\phi$ spectral function at vanishing $\phi$ momentum as a function of the $\phi$ energy $E_\phi$. The left panel is devoted to the case when considering the $\bar K K$ self-energy alone (dashed) or the cumulative contribution of both the $\bar K K$ cloud and $\phi N$ interactions in the HLS model (dash-dotted) and the $SU(6)$ model (solid), all the three cases considered at $\rho=\rho_0$. The $\phi$ spectral function in vacuum is also shown for reference (dotted).
The addition of new decay channels in nuclear matter when the $\bar K K$ cloud is modified, such as $\phi N\to K Y$ and $\phi N \to [K \Lambda(1405)]\to K \pi Y$, already induces a substantial broadening of the $\phi$ spectral function as compared to the vacuum case, as was already shown in Refs.~\cite{Klingl:1997tm,Oset:2000eg,Cabrera:2002hc}\footnote{In our former results from \cite{Cabrera:2002hc}, the $\phi$ width at normal matter density from the $\bar K K$ cloud was found to be of about $27$~MeV including vacuum decays.}, with a considerable population of the low-energy tail (which is further enhanced at high temperatures \cite{Cabrera:2015xra}) and a practically negligible mass shift. When considering, on top of the former, the self-energy generated by the $\phi N$ interaction, an overall increase of the broadening is observed whereas at the same time some structure appears in the spectral function. Notably, in the $SU(6)$ model, the $\phi$ quasi-particle peak is shifted to lower energies in agreement with the observed attractive optical potential in the previous section. Moreover, the spectral function develops a second shoulder above the $\phi$ mass which corresponds to the excitation of $N^*N^{-1}$ modes from the resonant $\phi N$ scattering in the $\gtrsim 2$~GeV region.
When using the self-energy generated by the HLS interaction, the effects around the $\phi$ mass are much milder, as anticipated in the discussion of the $\phi$ optical potential, whereas a similar shoulder appears in the $E_\phi=1100-1200$~MeV region of identical origin as in the $SU(6)$ calculation. In both approaches, the low-energy part of the $\phi$ spectral function is also enhanced with respect to the $\bar K K$ contribution. This is originated by the coupling to resonance-hole states generated by the vector-meson baryon interaction in the far sub-threshold region, such as the $N(1650)$ and other three-star candidates in the $1600-1700$~MeV interval.
The right panel of Fig.~\ref{fig:Spec-contrib} shows the evolution with the density of the $\phi$ spectral function from vacuum up to $\rho=1.5\rho_0$ in the $SU(6)$ evaluation of the $\phi N$ contribution, which produces the maximal effect within the uncertainties of the models discussed for the $\phi N$ interaction. 

\section{Summary and Outlook}
\label{sec:summary}
In summary, we have studied the role of resonant $\phi N$ interactions in the renormalization of the $\phi$-meson properties in a dense nuclear medium. We have calculated the $\phi$ self-energy by summing the $\phi N$ scattering amplitude over the allowed nucleon states in the Fermi gas approximation, which automatically incorporates the effect of Fermi motion. Additional medium corrections such as Pauli blocking on intermediate nucleon states have also been studied. 

Two effective models for vector-meson baryon interactions, based on hidden local gauge and $SU(6)$ spin-flavor symmetries, have been implemented to better assess theoretical uncertainties. Both approaches, which implement the unitarity of the $T$-matrix in coupled channels, generate resonant states in the vicinity of the $\phi N$ threshold with a strong $s$-wave coupling to $K^*Y$ intermediate states.
The $\phi$ optical potential exhibits a sizable dependence on the $\phi$ momentum, particularly below $1$~GeV$/c$, as a consequence of the resonant character of the $\phi N$ interaction. The self-energy induced by this novel mechanism is of the same order of magnitude as the contribution from in-medium $\phi\to \bar K K$ decays, effectively accounting for additional absorption channels as $\phi N\to K^* \Lambda, K^* \Sigma$.
Our full results are in reasonably good agreement with the low-density approximation ($\Pi_{\phi}\sim T_{\phi N}\rho$). Pauli blocking in the scattering amplitude has a more pronounced reflection in the self-energy at $\phi$ momenta below $600$~MeV$/c$.

Altogether ($\bar K K$ and $\phi N$), from the optical potential at zero momentum we obtain an in-medium $\phi$ decay width up to $\sim 70$~MeV at normal nuclear matter density in the $SU(6)$ approach, further increasing with momentum up to practically $100$~MeV in the range $400-500$~MeV$/c$, and slowly decreasing at higher momenta. Similarly, an attractive $\phi$ mass shift up to $\sim 35$~MeV can be extracted from the real part of the optical potential.
These effects are more moderate in the HLS approach to the $\phi N$ interaction, particularly at threshold, which calls for a more detailed study of these effective models. In view of recent (and future) experimental data, particularly the momentum binning analysis of the nuclear transparency ratio, the $\phi$ production reaction in nuclei could be used to better constrain the vector-meson baryon interaction at low energies.

The $\phi$ spectral function exhibits a large broadening as compared to the vacuum case, with a shift of the quasi-particle peak to lower energies, consistently with the attraction observed in the optical potential. When adding the resonant $\phi N$ interaction, a second bump develops above the nominal $\phi$ energy from the mixing of the $\phi$ with resonant-hole excitations. This feature is observed in the two model approaches to the $\phi N$ interaction, whereas the effects around the $\phi$ mass are much more moderate in the HLS approach. Additional far sub-threshold modes are populated in the spectral function on top of the ones accounted for from absorption via $\bar K K$ decays.

Our results point at a strong and energy dependent enhancement of $\phi$ absorption in the nuclear medium from resonant $\phi N$ interactions, with in-medium $\phi$ decay widths fairly in line with nuclear production experiments. A suitable analysis of the nuclear transparency ratio for different nuclei as a function of momentum is called for with our updated $\phi$ self-energy.
Our results, in terms of the $\phi$ nuclear optical potential and spectral function, are also suitable for applications in transport simulations of heavy-ion collisions, where the calculation of dynamical quantities requires that the pertinent reaction rates or transition probabilities are folded with the spectral functions of the particles in the initial and final states. Work along these lines within transport approaches such as the Parton-Hadron-String Dynamics and the Isospin Quantum Molecular Dynamics models is in progress.

\section*{Acknowledgements}
We acknowledge fruitful discussions with Elena Bratkovskaya, Wolfgang Cassing, Juan Nieves, Eulogio Oset, \`Angels Ramos and Laura Tol\'os. This research has been partially supported by the Spanish Ministerio de Econom\'{\i}a y Competitividad (MINECO) and the European fund for regional development (FEDER) under Contracts FIS2011-28853-C02-01, FIS2014-51948-C2-1-P, FIS2014-51948-C2-2-P and SEV-2014-0398, by the Helmholtz International Center for FAIR within the framework of the LOEWE program, and by Generalitat Valenciana under Contract PROMETEOII/2014/0068. A.N.~Hiller Blin acknowledges support from the Santiago Grisol\'{\i}a program of the Generalitat Valenciana. D.~Ca\-bre\-ra acknowledges support from the BMBF (Germany) under project No.~05P12RFFCQ.

\end{document}